\documentclass[journal,oneside]{IEEEtran}
\usepackage{times,amsmath,epsfig,amsfonts,amssymb,graphicx,url,cite,subfigure,colortbl,bm,upgreek,multirow,booktabs}

\begin{document}
\hyphenation{}
\title{Empirical Evaluation of a 28~GHz Antenna Array on a 5G Mobile Phone Using a Body Phantom}
\author{Lauri V\"{a}h\"{a}-Savo,~%~\IEEEmembership{Student~Member,~IEEE,}
	Christian Cziezerski,~%~\IEEEmembership{Student~Member,~IEEE,}
	Mikko Heino,~%~\IEEEmembership{Student~Member,~IEEE,}
        Katsuyuki Haneda,~\IEEEmembership{Member,~IEEE,} 
        Clemens Icheln,~%\IEEEmembership{Member,~IEEE,}
        Ali Hazmi and~%~\IEEEmembership{Student~Member,~IEEE,}
	Ruiyuan Tian~%\IEEEmembership{Member,~IEEE}
	
\thanks{L.  V\"{a}h\"{a}-Savo, C. Cziezerski, M. Heino, K. Haneda and C. Icheln are with the Department of Electronics and Nanoengineering, Aalto University-School of Electrical Engineering, Espoo FI-00076, Finland (e-mail: lauri.vaha-savo@aalto.fi).}% <-this % stops a space
\thanks{A. Hazmi and R. Tian are with Huawei Technologies Finland (e-mail: ali.hazmi@huawei.com).}% <-this % stops a space
\thanks{The work of M. Heino, K. Haneda and C. Icheln is in part supported by the Academy of Finland research project "Massive MIMO: Advanced Antennas, Systems and Signal Processing at mm-Waves (M3MIMO)", decision \#288670.}% <-this % stops a space

}% <-this % stops a space

\maketitle

\begin{abstract}
Implementation of an antenna array on a 5G mobile phone chassis is crucial in ensuring the radio link quality especially at millimeter-waves. However, we generally lack the ability to design antennas under practical operational conditions involving body effects of a mobile user in a repeatable manner. We developed numerical and physical phantoms of a human body for evaluation of mobile handset antennas at $28$~GHz. While the numerical human model retains a realistic and accurate body shape, our physical phantom has much simpler hexagonal cross-section to represent a body. Gains of the phased antenna array configuration on a mobile phone chassis, called co-located array is numerically and experimentally evaluated. The array is formed by placing two sets of $4$-element dual-polarized patch antenna arrays, called two modules, at two locations of a mobile phone chassis. Modules are intended to collect the maximum amount of energy to the single transceiver chain. Spherical coverage of the realized gain by the array shows that the experimental statistics of the realized gains across entire solid angles agree with numerical simulations. We thereby demonstrate that our antenna evaluation method reproduces the reality and our phantom serves repeatable tests of antenna array prototypes at $28$~GHz.
\end{abstract}

\section{Introduction}
The fifth-generation cellular networks, which are rolled-out at the moment, promise higher data rates compared to the legacy systems. Using radio frequencies above 6 GHz is one of the new ways to achieve the promised higher data rates. These radio frequency bands are called new radio frequency range 2 (NR FR2) in addition to legacy below-6 GHz radio frequencies called NR FR1~\cite{TR38810}. Despite the increasing relevance of NR FR2, the effects of a human body on the gains and losses to the antennas have got relatively little attention at NR FR2. 
Despite the significant studies ~\cite{Halender16_AWPL, Syrytsin17_TAP, Zhao17_AWPL, Hejselbaek17_TAP, Zhao17_TAP, Hong17_TAP, Yu18_TAP, Haneda18_VTCS, Haneda18_EuCNC, Xu18_IEEEAccess, Syrytsin18_TAP, Raghavan19_ComMag, Raghavan19_TC, Hazmi19_EuCAP,  Zhao19_IEEEAccess, Heino19_EuCAP, Cziezerski19_APC} that have given insights into robust mobile phone antenna array configurations at NR FR2 on a realistic environment involving human body effects, clear design guidelines for mobile antenna systems are still missing.

When consider the empirical evaluation of manufactured antenna elements and arrays at NR FR2, the works are even fewer because of challenges in performing the necessary measurements. One of the major obstacles in performing an empirical evaluation of antennas under the influence of the mobile phone user is the lack of a human body phantoms. All works reporting empirical evaluation of antenna radiation or radio-link characteristics with a mobile user~\cite{Syrytsin17_TAP, Zhao17_AWPL, Hejselbaek17_TAP, Zhao17_TAP, Syrytsin18_IEEEAccess, Zhao19_IEEEAccess, Raghavan19_ComMag} use the real human subjects. It is known that measured antenna radiation characteristics with a real human subject are not repeatable.

We therefore in this paper report our development of a human body phantom aiming at repeatable measurements of antenna's radiation characteristics. Earlier works on human body phantoms~\cite{Gustafson12_RE, Aminzadeh14_EL, Lacik16_AWPL} are not intended for the evaluation of mobile phone antennas in their proximity, though they may well be usable for the purpose. Repeatable measurements with the developed phantom enable us to compare the performance of different antenna array configurations and especially to identify array configurations that are robust against the electromagnetic interaction with the human body including shadowing. Repeatable antenna measurements under human influence are also essential in device conformance test. Having integrated the developed phantom into our far-field antenna measurement facility, we empirically evaluate the gains of a phased antenna array implemented on a mobile phone chassis with a single transceiver chain. Antenna array was chosen with the goal to provide generic but realistic-enough example for demonstrating the use of developed phantom. Our exemplary array consists of two modules of patch antenna arrays installed at different sides of a mobile phone chassis, which we reported in~\cite{Cziezerski19_APC} as a co-located array. The array allows us to collect multipath powers arriving from different directions. In this present paper, parts of the reported insights in~\cite{Cziezerski19_APC} are empirically verified. We aim at reproducing scattered fields from a realistic human body using a physical body phantom that has a simple shape. To this end, we compare the experimental results using the simple body phantom to the simulations with realistic numerical human model. We study spherical coverage of the realized antenna gains. E.g.,~\cite{Halender16_AWPL} uses the spherical coverage to estimate coverage efficiency and nowadays it is introduced by the 3GPP~\cite{TR38810} as a metric to evaluate and rank phased antenna arrays. 
\begin{figure*}[ht!]
	\begin{center}
		\includegraphics[scale=0.43]{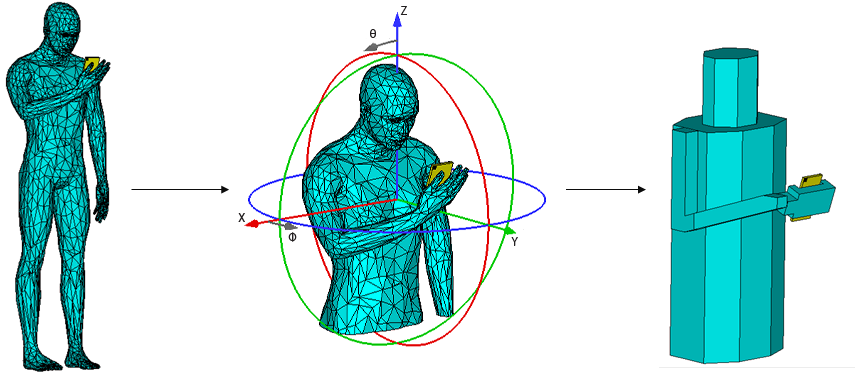}
		\caption{A realistic numerical human body, truncated numerical human body in coordinated system used in this paper and hexagonal body phantom.}
		\label{fig:accurate_human}
	\end{center}
\end{figure*}

In summary, novel contributions of the present paper are summarized in three-fold as
\begin{enumerate}
 \item We develop {\it a physical human body phantom} for repeatable experimental evaluation of electromagnetic interaction between human body and mobile handset antennas at $28$~GHz, which is essential for comparison of array designs and device conformance test;

 \item We perform experimental study of antenna radiation under the presence of the developed human body phantom, along with its comparisons with a numerical study; and finally,
 
 \item We confirm feasibility of repeatable antenna measurements using the developed human body phantom, revealing their uncertainty in terms of realized gains.
\end{enumerate}

The rest of the paper is organized as follows: in Section~II, we use a numerical human body model to identify influential body parts that affect radiation of antenna elements on a mobile phone in browsing mode. In the same section, we then elaborate the proposed physical human body phantom. In Section~III we introduce the antenna element, antenna array modules and their placement on a mobile phone to realize co-located array. Feed line designs and radiation characteristics of the prototype antennas are also shown. In section~IV, we show the measured scattered fields from a single-element antenna, defines the pattern synthesis method and spherical coverage. Simulated gains are compared with measurements for validation of our scientific approaches. Finally, we summarize the main conclusions in Section~V.

\section{Human Body Phantoms}
\label{sec:phantom}
In this section, numerical and physical phantoms are introduced. The numerical phantom represents realistic human body shape and is introduced to identify body parts that we need to consider in the physical phantom. After the numerical phantom is introduced, we create skin material and place it on top of a hexagonal cross-sectioned body phantom. 

\subsection{Numerical Phantom}
\label{sec:Numphantom}

A 3D model of a realistic numerical human body, shown in Fig.~\ref{fig:accurate_human}, was created by the open-source tool Make Human~\cite{MakeHuman}. The human model was imported to CST Microwave Studio where surface impedance representing skin was added as an attribute of the model. Because of the small penetration depth of $0.92-0.95$~mm~\cite{Gabriel97_PMB} at $28$~GHz, the surface model allows us to calculate scattered fields of antennas accurately with FTDT simulations. The integral equation solver in CST Microwave Studio was used to calculate the scattered fields~\cite{Heino19_EuCAP}. The simulation covered the case when a human holds a mobile phone in portrait browsing mode using a single hand. In our study, fingers and hand palm do not touch any antenna elements, avoiding the case of reducing the impedance matching, which is not in the scope of this study. First, we derive equivalent near-field sources for a cuboid sub-volume enclosing only the hand and the mobile phone, with one feed activated at a time, hence we obtain sixteen equivalent sources using FTDT simulations. These equivalent field sources are used to compute the radiated fields with the body model using the surface integral equation solver. The use of equivalent field sources allows us to simulate the antenna feeds and body separately, which usually is necessary due to the very different scales of mesh sizes required. When solving the integral equations with the equivalent near-field sources of antenna elements, the mobile phone chassis is included inside the equivalent-source volume, but as a simple cuboid as a cuboid made of a perfect electric conductor.

\begin{figure}[ht]
	\begin{center}
		\includegraphics[scale=0.4]{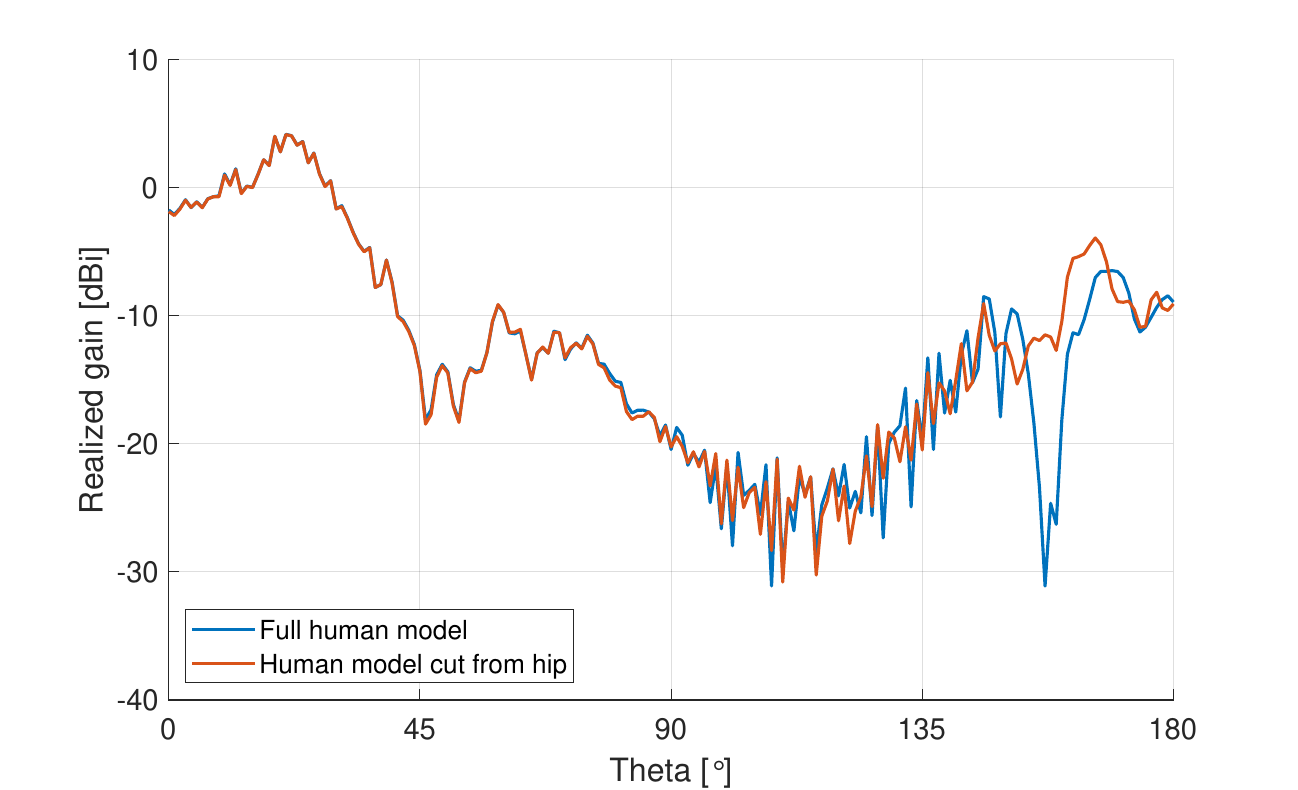}
		\caption{Comparison of realized gain of an antenna element, simulated with truncated and full-sized numerical human models; the gain patterns represent $\phi = 270^\circ$ cut.}
		\label{fig:MAE}
	\end{center}
\end{figure}

The numerical human model was truncated just below the hip level to reduce the complexity of simulation and measurement. By cutting the human model just below the hip, the mean absolute error for the realized gain for each constant $\theta$ pattern cut was below $0.3$~dB when $\theta < 107^\circ$ and below $0.7$~dB when $\theta < 140^\circ$ compared to simulations with the full-sized human model. Figure~\ref{fig:MAE} shows the difference of the realized gain simulated with the truncated and full-sized human models. When a human holds a mobile phone, field radiation to legs, e.g., for $\theta \geq 150^\circ$, is usually least important in cellular radios as radiated fields bounce from the ground, reflect back to the sky and never reach a base station. Therefore, the truncated human model was considered sufficiently accurate for evaluation of cellular mobile radios.

\subsection{Physical Phantom}
\label{sec:phys_phantom}
The thickness of human skin varies between $1.3$ and $2.9$~mm \cite{Wu15_MM}, and is thicker than the penetration depth of electromagnetic waves at $28$~GHz \cite{Gabriel97_PMB}. We therefore need to consider only skin in our phantom design. Gabriel's extrapolated results have been experimentally verified at higher frequencies including $28$~GHz \cite{Sasaki14}. Chahat et al. created skin material at $60$~GHz~\cite{Chahat12_EL} and the similar material was shown to work also at lower frequencies~\cite{Dancila14_SRT}. This skin phantom material consists of deionized water, agar, polyethylene-powder and TX-151. Water is the main constituent of the material since human skin is mostly composed of water. Water determines the dispersive behavior of the material. Agar is used for the shape retention, and its effect to the dielectric properties is minimal. Polyethylene-powder is used to decrease the real part and adjust the imaginary part of the permittivity. TX-151 is used to increase the viscosity of the material since agar and polyethylene-powder cannot be mixed directly. Skin mimicking material containing $70$~m\% of water was manufactured according to Table~\ref{tab:recipe}.

\begin{table}[htb]
\begin{center}
\caption{Materials and amounts used for phantom mimicking human skin} 
\label{tab:recipe}
\begin{tabular}{|c|c|}
\hline
Ingredients & Mass (g) \\ \hline
Deionized water & 100 \\ \hline
Agar & 3  \\ \hline
TX-151 & 4 \\ \hline
Polyethylene powder & 36  \\ \hline
\end{tabular}
\end{center}
\end{table}

The permittivity of the skin material was measured using a well-established method to observe transmission coefficients of the material elaborated in \cite{Olkkonen13_EuCAP}. Table~\ref{tab:Permittivity} shows the real part of permittivity and conductivity of the skin material at $28$~GHz. The real part is 3.3\% less and the conductivity is 4.6\% more than skin in Gabriel’s model. The permittivity measurements were repeated monthly for four months to determine its durability. The skin material sample was kept in an airtight container between measurements to refrain water from evaporating. Changes in the real and imaginary parts of the permittivity were noticeable only after four months. Thus, this skin material can be used up to three months after the manufacturing if kept in an airtight environment.

\begin{table}[htb]
\begin{center}
\caption{The permittivity of skin material at 28 GHz} 
\label{tab:Permittivity}
	\begin{tabular}{|c|c|c|}
\hline
 & Real part & Conductivity \\ 
 & of permittivity & \\\hline
Gabriel's model & 16.55 & 25.82\\ \hline
Measurement & 16 & 27 \\ \hline
\end{tabular}
\end{center}
\end{table}

The skin material is cast onto flat plates and left for about 12 hours. Then the material was semi-solid\footnote{Semisolid in our paper means that the material is solid and flexible when it is in the room temperature and liquid when cast on the plate. Increasing the amount of TX-151 and agar makes the material harder.} and could be attached to the surface of the Styrofoam phantom core
. This plastic film is made of $0.1$~mm thick LD-PE, which has a relative permittivity of $2.3$ \cite{Krupka16_MWCL}. Reflection coefficient of the film was $-28.4$ dB at $28$~GHz according to ITU-R P.2040~\cite{ITU-R_P2040}, which indicated that the film has no effect on the reflection coefficient of the skin material. 

It is not easy to manufacture a skin material layer as thin as $2$~mm. To ensure structural durability of skin layer, its thickness on the phantom was set to $5$~mm. According to calculation from the ITU-R P.2040, the reflection coefficient of a skin layer with $5$~mm thickness differs only $0.14$~dB compared to skin with $1.5$~mm thickness. So, we can expect negligible effects on electromagnetic wave scattering and reflection while ensuring durability of the phantom.

\begin{figure}[!ht]
	\begin{center}
		\includegraphics[scale=0.48]{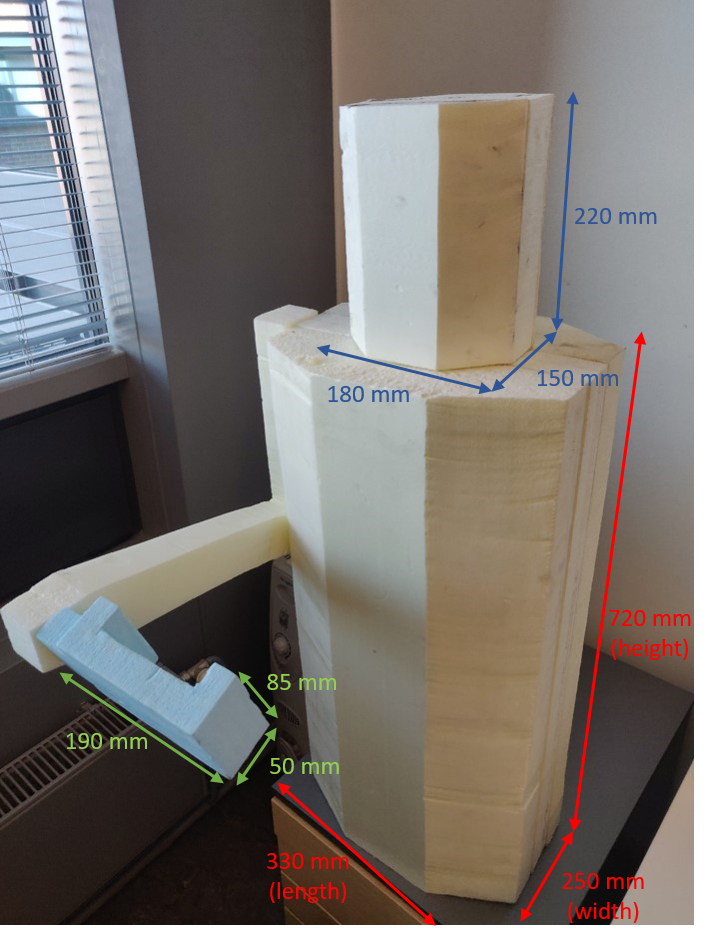}
		\caption{The Styrofoam structure of hexagonal body phantom without skin material.}
		\label{fig:phantom}
	\end{center}
\end{figure}

The numerical human model shown in \cite{Heino19_EuCAP} and \cite{mikko_heino_2019_3249975} was used as basis for creating a simplified physical human phantom. We take the diameters, the pose and the phone tilt angle from the numerical human model but the shapes needs to be simplified to ensure manufacturability of the phantom. The analysis of the near-field effects that the hand would have on antenna characteristics is out of the scope of this paper. Therefore, the designed handgrip ensures a separation of wavelength between phantom hand and antenna array. The phantom consists of hexagonal pillar shaped head and torso and rectangular arm and hand palm pieces. The hand palm is a rectangular box with an immersion for the mobile phone chassis equipped with antenna array. The base material of the inner parts of the phantom is Styrofoam, and a wooden pole in the center connects the head to the torso. Styrofoam is chosen because it is light, unyielding and is relatively easy to shape. The skin material is added on all surfaces of the Styrofoam structure except for the bottom where the mechanical connection to the measurement tower is implemented. Dimensions of the torso and head of the phantom are $330 \times 250 \times 720$~mm and $180 \times 150 \times 220$~mm (length, width, height), respectively. The hand palm is a rectangular box with dimensions $190 \times 50 \times 85$~mm and the immersion is $100 \times 20 \times 85$~mm. The arm is made from a $430 \times 300 \times 100$~mm rectangular piece of Styrofoam. The palm is attached to the arm with a $20^\circ$ tilt angle from the arm orientation, as illustrated in Fig.~\ref{fig:phantom} with dimensions. The tilt ensures a realistic position of the phone with respect to the head. The 3D model of this Styrofoam structure of hexagonal body phantom can be found from \cite{vaha_savo_lauri_2021_4558779}.

\section{Co-Located Phased Antenna Array on a Mobile Phone Chassis}
\subsection{Antenna Element}
\begin{figure}[t]
	\begin{center}
		\subfigure[]{\includegraphics[scale=0.3]{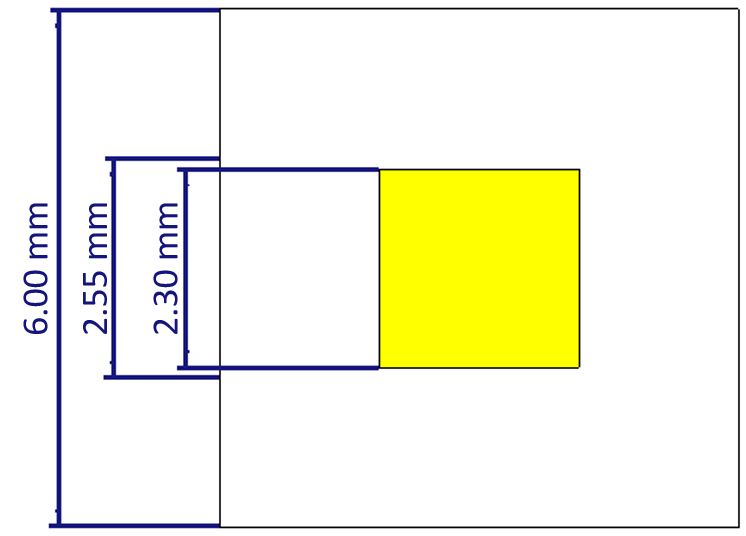}
			\label{fig:antenna_element_top}}
		\subfigure[]{\includegraphics[scale=0.17]{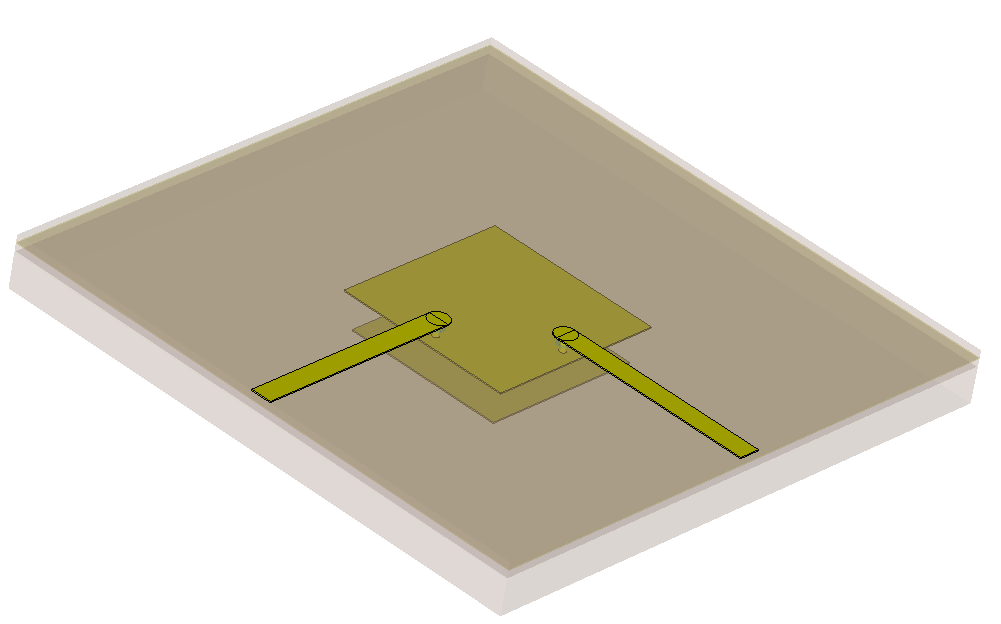}
			\label{fig:stacked_patch}}
		\caption{(a) Dimension of a single stacked patch antenna element, top view. (b) Dual-polarized microstrip feed structure of a single stacked patch element, bottom view with transparent ground plane and PCB. 
		}
		\label{fig:antenna_element}
	\end{center}
\end{figure}

In this work we decided to use a rectangular stacked patch as an antenna element intended for a mobile phone, which is detailed in~\cite{Cziezerski19_APC}. The patch is designed so that it realizes a wide enough bandwidth for radio systems operating around $28$~GHz. The Relative permittivity and dissipation factor of the substrate is $3.6$ and $0.004$, respectively, which are available from Rogers as RO4450B with $0.505$ and $0.101$~mm thickness. Horizontal dimensions of the antenna element are shown in Fig.~\ref{fig:antenna_element}(a) and the feed structure on the backside is illustrated in Fig.~\ref{fig:antenna_element}(b). Two microstrip feed lines are implemented on $0.101$~mm-thick RO4450B substrate on the opposite side of the ground plane of the patch antennas. The total thickness of the antenna integrated with the feed lines is $0.775$~mm. Feed lines are galvanically connected to the patch through microvias. Two feeds are needed for each patch antenna element as it covers the two orthogonal polarizations.

\subsection{Co-located array}%Antenna Array}
\begin{figure}[t]
	\begin{center}
    \includegraphics[scale=0.3]{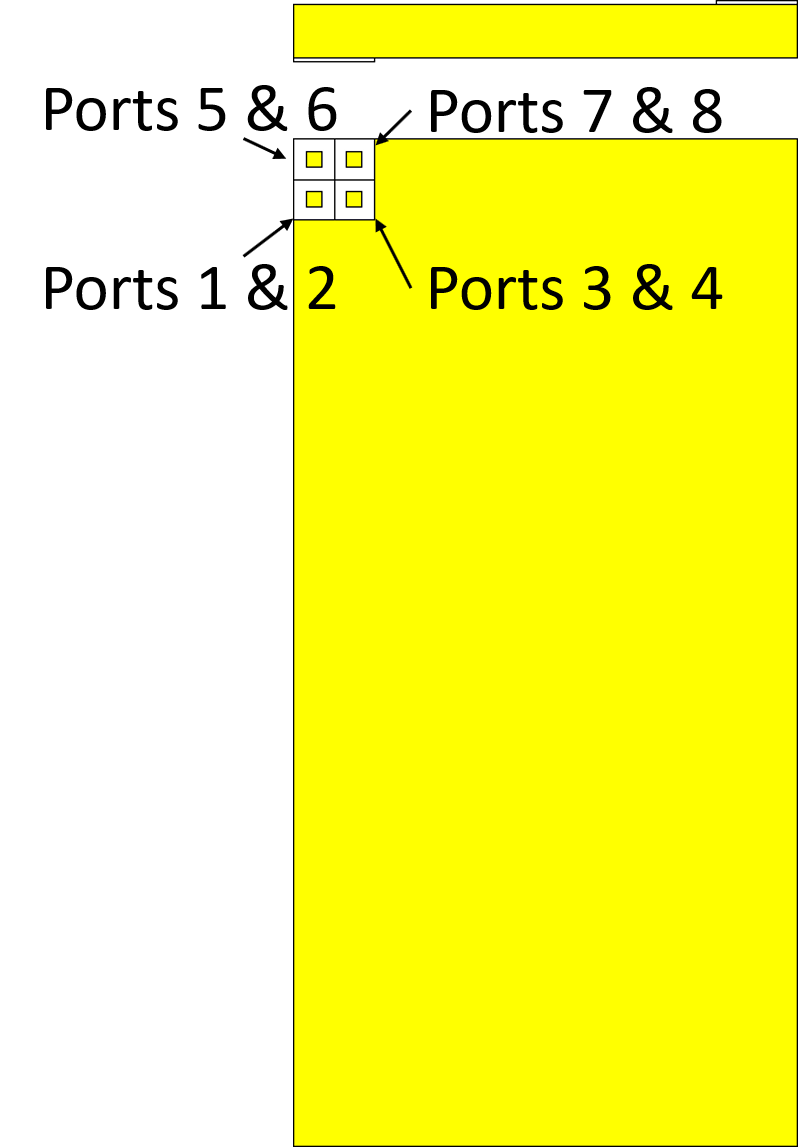}
		\caption{The co-located array configurations, seen from front and top sides of the mobile phone chassis with one module on front- and one on back-side. Used in simulations (two discrete ports per element adding up to 16 ports in total)}
		\label{fig:array_geometry}
	\end{center}
\end{figure}

\label{sec:array}
The dual-polarized stacked patch antenna element in Fig.~\ref{fig:antenna_element} is used to form a $2\times 2$ square array. The array is called a {\it planar array module} hereinafter and has it $8$ feeds. Figure~\ref{fig:array_geometry} illustrates the {\it Co-located array} where one planar modular array is at the top-left corner on the front side of a mobile phone and the second one at the top-right corner on the back side. These two modules are designed to cover different parts of the space and collect the as much energy to the transceiver since the patch antenna will not radiate to the backside.

\begin{figure*}[ht!]
	\begin{center}
		\includegraphics[scale=0.45]{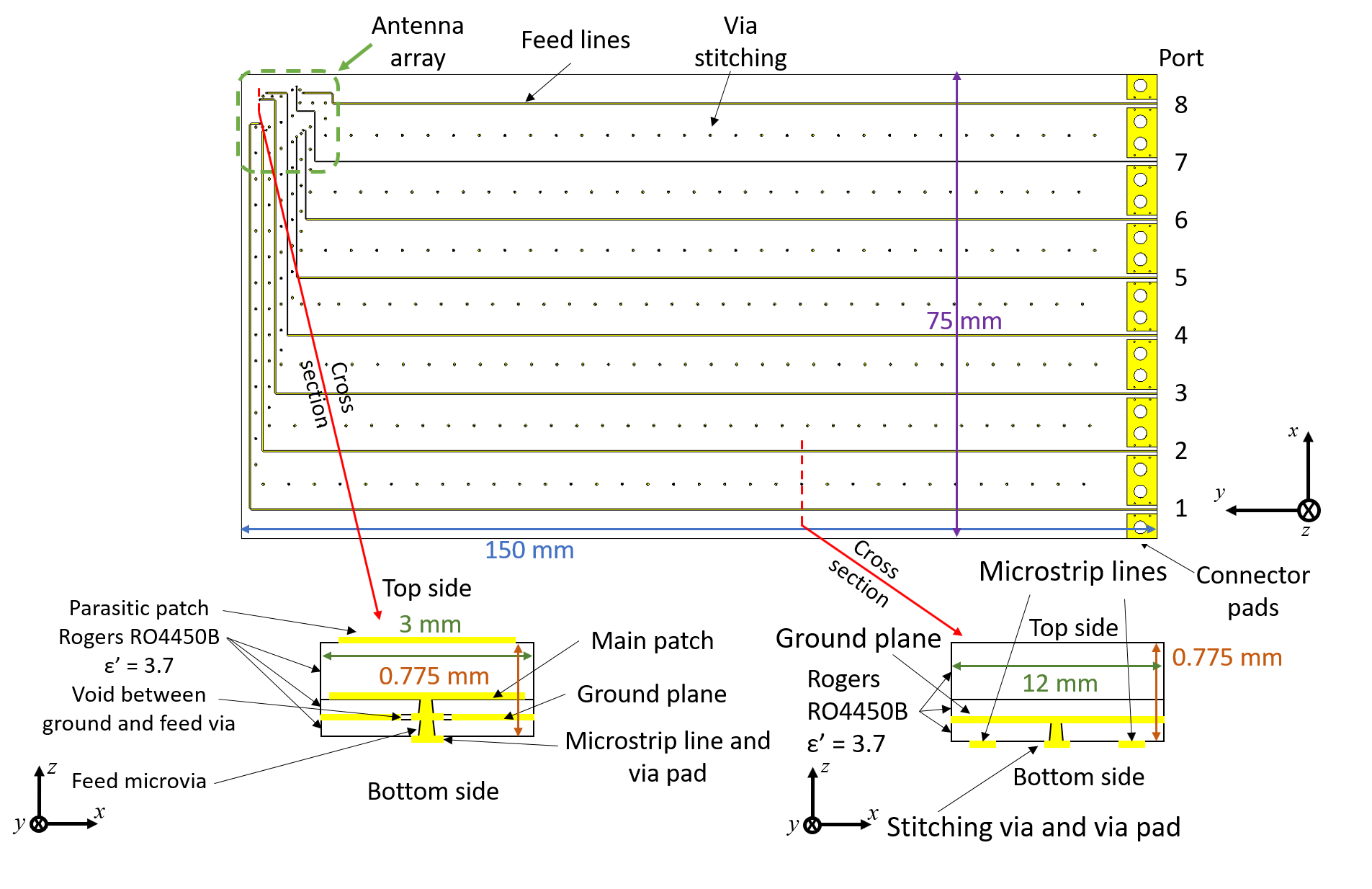}
		\caption{Bottom view of the PCB including microstrip feed line structure and two cut planes showing the stackup of the PCB. Yellow and white parts represent the metal and substrate.} 
		\label{fig:array_feed_network}
	\end{center}
\end{figure*}
\subsection{Fabrication of the Array}
\label{sec:fabrication}

\begin{figure}[t]
	\begin{center}
		\includegraphics[scale=0.40]{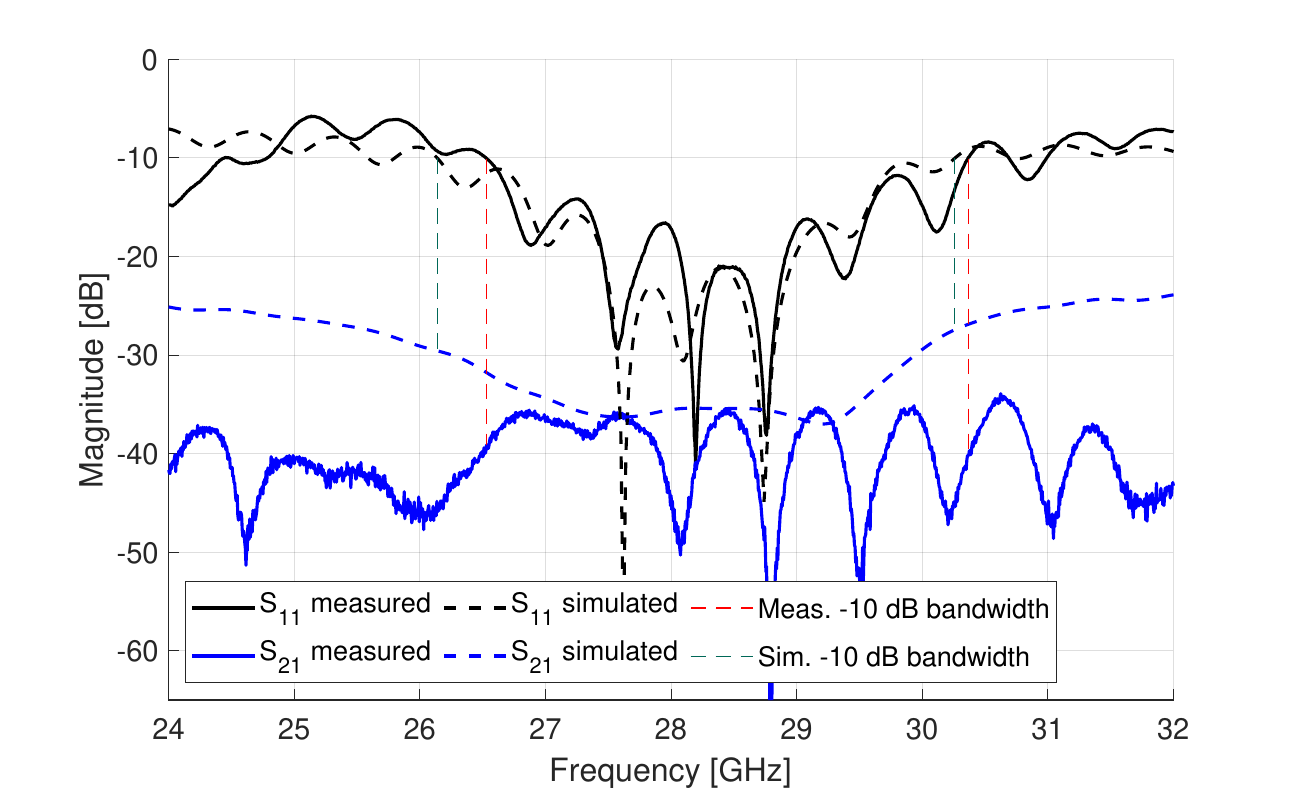}
			
		\caption{Reflection coefficient and mutual coupling of a single port.}
		\label{fig:antenna_matching}
	\end{center}
\end{figure}

The array is placed on a phone-sized structure, with the dimensions of $150\times 75 \times 8$~mm. The printed circuit board (PCB) was designed to be $150\times 75$~mm. We want to minimize the possible effect of the connectors on the radiation of the array. We line up all eight connectors at the bottom side of the phone chassis and design the microstrip feed lines between the connectors at the bottom and the antenna elements at the top corner of the PCB. As the front and back sides of the antenna array in the phone chassis are symmetrical, we manufactured only one side of the phone and in the measurements, we flipped the chassis around to realize the other module. Having only one module on the phone chassis ensures enough space for the eight connectors. The feed line width is $0.208$~mm. The feeds of the stacked patch antennas were manufactured as laser-drilled microvias with a diameter of $0.125$~mm and via pads with a $0.275$~mm diameter, as indicated in the bottom-left cross-section schematic of Fig.~\ref{fig:array_feed_network}. Using microvias instead of normal-sized vias ensures minimal parasitic capacitance and inductance for antennas. A void ring of a $0.375$~mm diameter was introduced around the microvia to avoid galvanic connection between ground and feed via. On the bottom-right cross-section drawing of Fig.~\ref{fig:array_feed_network} , via pads are introduced on the bottom side of the PCB and are stitched to the ground plane through microvias. The stitching is introduced between parallel microstrip line pairs to decrease the coupling. Dimensions of via pads and microvias are the same as those of antenna feeds. Since the PCB is only $0.775$~mm thick, we added $5$~mm thick Rohacell-foam and $1.5$~mm thick FR4-substrate on the bottom side of the PCB. These additions strengthen mechanical stability of the PCB and making the total thickness similar to that of a mobile phone. The ground plane of the manufactured PCBs was connected to FR4 substrate by a copper tape to ensure no radiation would leak from inside of the mobile phone, especially from connector pins. The array uses Southwest Microwave’s narrow-block $2.40$~mm end-launch connector (Mfr. No: $1492$-$04$\rm{A}-$9$). The feed line structure of the array is shown in Fig.~\ref{fig:array_feed_network}. All the feed lines have different lengths and this introduces different losses. The simulated losses caused by the feed lines are between $4.3$ to $6.5$~dB. The measured and simulated reflection coefficient of one representative port as well as the mutual coupling between two ports are illustrated in Fig.~\ref{fig:antenna_matching}, showing $3.8$ and $4.1$~GHz of bandwidth for measured and simulated reflection coefficient for $-10$~dB matching levels, respectively. They correspond to $14$~\% and $15$~\% of relative bandwidth centered at $28$~GHz, respectively. The mutual coupling was as low as $29$~dB for measured and $27$~dB for simulated case across the $-10$~dB impedance bandwidth of the antenna. The oscillations seen in the reflection coefficient are caused mainly by impedance mismatch at the $90^\circ$ turns in the feed lines.

\section{Antenna Gain Evaluation}
Having set up numerical models of the antenna array and hexagonal body phantom and having fabricated them, we now evaluate antenna gains from simulations and measurements and then compare them to verify our models and methods. Radiated far-fields of each antenna feed in the module-based antenna array, is first simulated and measured in free space in order to estimate the losses of all feed lines and connectors in the fabricated antenna array. We can de-embed these losses from the far-field radiation patterns measured with the hexagonal body phantom. Then, ideal pattern synthesis is applied to the element far-field patterns to analyze the spherical coverage of the realized gain of the array without and with the phantom. Efficacy of our approach is evaluated by comparing the spherical coverage statistics of realized gains by the array between simulations and measurements. All simulated and measured patterns are available in \cite{vaha_savo_lauri_2021_4558779}

\subsection{De-embedding Feed Lines and Connector Losses}

In the simulations, we can feed the antenna elements directly at the two feed pins of each patch. However, in measurements we need feed lines along with connectors and patch cables, for which neither direct measurements nor simulations provide reasonable loss estimates, since e.g. hand soldering introduces different losses to all connectors. We therefore estimate the losses by calculating the mean difference between simulated and measured main beams for each individual feed port in free space. The main beams span across $\pm 60^\circ$ in both $\theta$ and $\phi$ direction. The estimated losses have a form of complex amplitude and are later applied to the measured beam patterns with the hexagonal body phantom. Phase correction can be neglected because for spherical coverage analysis we defined antenna weights by ideal three-bit phase shifters, leading to $512$ random different phase combination patterns for each sub-array. From those $512$ phase combination patterns, we will find the best phase combination without phase de-embedding. Table~\ref{deembed} shows implementation-loss estimates for the manufactured array where port numbers are same as in Fig. \ref{fig:array_geometry}. The loss estimates are proportional to the length of the feed lines.

\begin{table}[htb]
\begin{center}
\caption{Loss estimates of cables, connectors and feed lines for each port of the array}
\label{deembed}
\begin{tabular}{|c|c|}
\hline

 & [dB] \\ \hline
Port 1 & 10.7  \\ \hline
Port 2 & 10.4  \\ \hline
Port 3 & 10.3  \\ \hline
Port 4 & 11.0  \\ \hline
Port 5 & 12.4  \\ \hline
Port 6 & 12.1 \\ \hline
Port 7 & 13.1 \\ \hline
Port 8 & 14.4 \\ \hline
\end{tabular}
\end{center}
\end{table}

\begin{figure}[t]
	\begin{center}
		\includegraphics[scale=0.5]{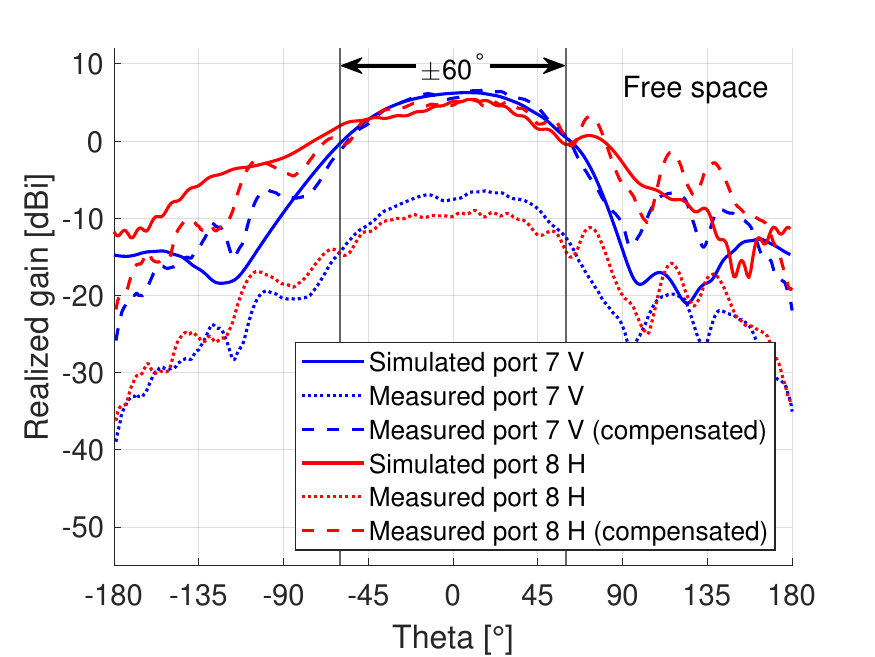}
		\caption{Azimuth cuts of the realized gain of array ports 7 V and 8 H ports in free space.}
		\label{fig:realized_gain_free}
	\end{center}
\end{figure}

Figure~\ref{fig:realized_gain_free} shows azimuth cuts of the realized gain patterns ports 7 and 8. Dotted lines are measured patterns, dashed lines are measured patterns with the de-embedding terms added and solid lines are simulated patterns. As can be seen, the compensated patterns agree well with the simulated ones in the main beam direction i.e. $\theta=\pm 60^\circ$. Maximum Gains of the shown measured ports are $6.3$~dBi and $5.3$~dBi and simulated maximum gains are $6.3$~dBi and $5.4$~dBi for port 7 and 8 in Fig. \ref{fig:array_geometry}, respectively. Back lobe levels are, however, slightly higher in compensated patterns than in simulations. This difference is probably caused by simplifications done to array models in order to get feasible simulation time. The exact cause of the difference is out of the scope of this paper. However, this increase in back lobe patterns does not affect to the result of this paper, namely the maximum realized gain in Fig. \ref{fig:realized_gain} and the CDF of that.  

\subsection{Pattern synthesis}
We have now obtained the simulated and measured polarimetric radiated far-field patterns separately for each feed, influenced by scattering from the human body. Implementation of phase shifting parts in pattern synthesis, whether it is a fully digital, analog or hybrid, is out of the scope of this paper. We assumed ideal, lossless phase shifting to concentrate on antenna-body interaction analysis. We calculate the weight for ideal pattern synthesis using the maximum ratio combining (MRC) which is known to be optimal weighting strategy in a single-chain receiver \cite{Molisch10_book}. The goal is to maximize the signal-to-noise ratio at the receiver rather than  forming clear main beams. The pattern synthesis is performed in the following manner. 
\begin{enumerate}
	\item We synthesize patterns of {\it four} antenna feeds out of 16 available feeds in the array. Four {\it sub-arrays} for pattern synthesis are defined in array for the purpose. One sub-array is a group of four patch antenna feeds radiating horizontally-polarized fields in a {\it planar array} (see Fig.~\ref{fig:array_geometry}), while the other is the group of four feeds for vertical polarization.
	\item The weights for synthesizing patterns applied at each antenna feed are defined by ideal three-bit phase shifters, leading to $512$ different patterns for each sub-array. As there are 4 sub-arrays, the total number of synthesized pattern realizations for an array configuration is $2048$. The $2048$ synthesized realizations of patterns overlap in some directions. We consider all of them relevant for the operation of the antenna array in practical operational environments. Not only synthesized patterns showing clear main beams, but also those without main beams are required to approximate the MRC. As examples, six out of $2048$ synthesized patterns are shown in Fig.~\ref{fig:pattern_realizations} for $\phi = 90^\circ$ cut. The first four mentioned patterns contribute to the maximum gain on this cut plane, while the last two beams overlap with others on this cut plane. We use one of the $2048$ realizations at a time for spherical coverage evaluation of the realized gains.
	 
	\begin{figure}[ht]
	\begin{center}
		\includegraphics[scale=0.5]{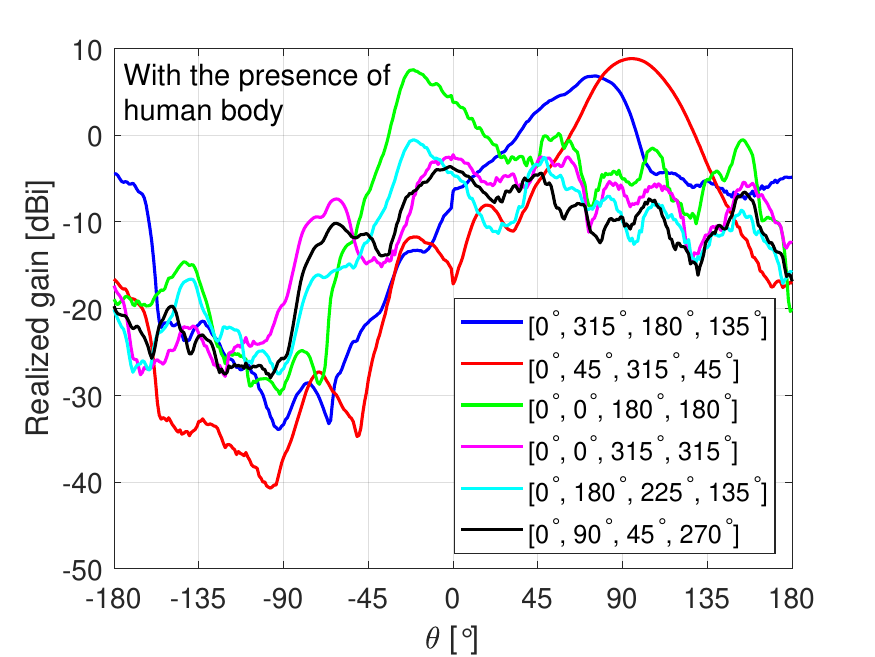}
		\caption{Six measured synthesized patterns out of $2048$ $\phi=90^\circ$ cut, under the presence of numerical human model. Numbers in the square bracket are phases given to different antenna elements by the ideal, lossless phase shifting.}
		\label{fig:pattern_realizations}
	\end{center}
    \end{figure}
\end{enumerate}

\subsection{Maximum Realized Gains and Spherical Coverage}
Spherical coverage is the empirical statistics of maximum gains that an antenna array can realize for all possible angles on a sphere. For an angle of interest $\boldsymbol{\Omega}$, the maximum realized gain of the array is defined by
\begin{equation}
\hat{G}(\boldsymbol{\Omega}) = \max_k G_k(\boldsymbol{\Omega}),
\end{equation}
where $G_k(\boldsymbol{\Omega}) = ||\boldsymbol{g}_k(\boldsymbol{\Omega})||^2$ is the power gain of $k$-th synthesized patterns of an array, $1 \le k \le 2048$ in our case; $\boldsymbol{g}_k(\boldsymbol{\Omega}) = [g_{\theta}(\boldsymbol{\Omega})~~g_{\phi}(\boldsymbol{\Omega})]$ is the complex gain vector of two field polarizations. Collection of $\hat{G}$ over different angles $\boldsymbol{\Omega}$ are characterized by its cumulative distribution function (CDF) for example as $CDF(x) = {\rm prob}(\hat{G} < x)$ where $\rm prob(\cdot)$ derives a probability of the condition specified in $(\cdot)$. 

When deriving the spherical coverage statistics, the angles $\boldsymbol{\Omega}$ are chosen to be uniform across the full sphere, so that the number of azimuth angles are smaller at higher elevation angles close to the pole than at lower elevation angles close to the horizon of the sphere. The uniform grid over the full sphere makes sure that the resulting spherical coverage is independent of the orientation of the mobile phone \cite{R4-1700095}.

\subsection{Measurement Setup and Error Estimation}

\begin{figure}[!ht]
	\begin{center}
	    \subfigure[]{\includegraphics[scale=0.33]{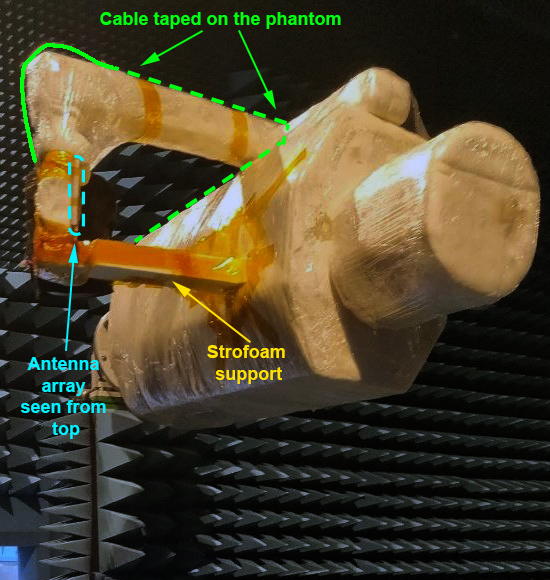}}\\
		\subfigure[]{\includegraphics[scale=0.38]{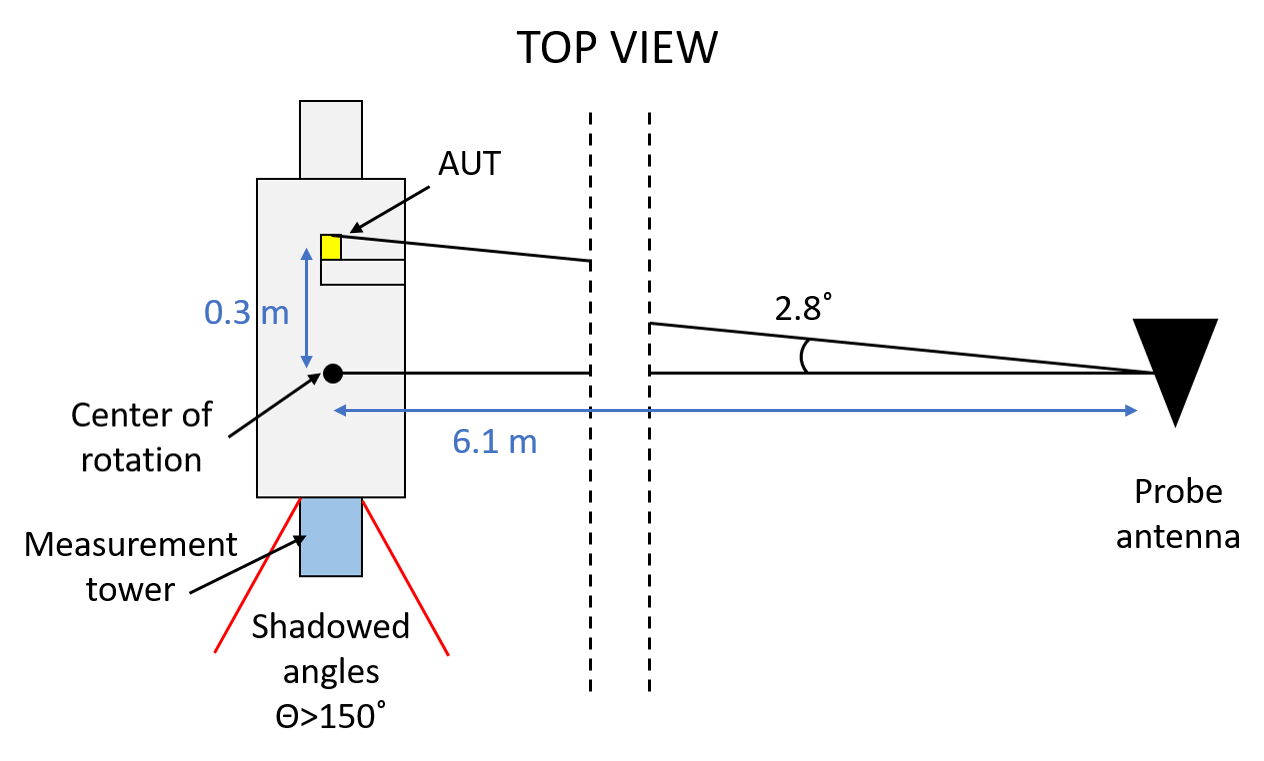}}
		\caption{(a) Picture of the hexagonal body phantom inside of the anechoic chamber. Solid and dashed green lines highlight the cable path where the cable is visible and cannot be seen in the picture, respectively. (b) Illustration of phantom attached to the measurement tower in an anechoic chamber.}
		\label{fig:meas_setup}
	\end{center}
\end{figure}

The measurements in free space and with the hexagonal body phantom were performed in the fully anechoic chamber in Aalto University built by ASYSOL. The measurement setup in the chamber consists of a VNA, a probe antenna, and two antenna towers, one for the probe antenna and the other for the AUT. The AUT can be rotated around two axes such that the full 3-D radiation pattern is recorded in one measurement session. The separation between the two towers is 6 m. The phantom is attached from its bottom end to the measurement tower, as shown in Fig.~\ref{fig:meas_setup}, and the phantom is therefore horizontally oriented with respect to the ground of the fully anechoic chamber. This way the shadowing caused by the tower happens in the direction of the legs, which is of relatively less importance in cellular radios as explained in in Section \ref{sec:Numphantom}. As indicated in the Fig.~\ref{fig:meas_setup}(a) a single RF cable from the antenna array is arranged to go behind the palm, then arm, and the side of the torso before it is connected to a port at the measurement tower. The cable was in the shadow of the phantom so it would not affect the radiation. The cable was 3 m long and its losses were calibrated in reference to a free-space measurement with the standard gain horn. Figure.~\ref{fig:meas_setup}(a) also indicates additional Styrofoam supports between e.g., the hand palm and chest to fix the arm properly during the measurements.

In the measurements we decided to measure with $1^\circ$ steps in $\theta$ direction and $10^\circ$ steps in $\phi$ direction. This way we ensure a realistic duration for our measurements. According to simulations, we do not lose accuracy in CDF of spherical coverage as compared to $1^\circ$-step also in $\phi$ direction. 

In the simulations, the skin permittivity is $\epsilon_r  = 16.55$ and conductivity $\sigma = 25.82$~S/m but when we manufacture the skin material we cannot fully control the dielectric properties. Here, we consider a variation of up to $\pm 10\%$ both in the real part of permittivity and in the conductivity, leading to eight additional cases that we compare to the original case without variation. 10\% variation is larger than the realized errors from our manufacturing indicated in Section II-B, but is considered as the worst-case scenario in other realizations of the skin material. We simulated their effects on spherical coverage statistics by evaluating the average difference of the CDF. From all eight cases with variations of permittivity and conductivity, we see an average difference up to $0.24$~dB with a standard deviation of $0.21$~dB. We can conclude that the effect of variations in material properties on the pattern and its statistics is minor. Another uncertainty source in measurements is the placement of the mobile phone in the rectangular hand palm phantom attached to the body.  While the intended distance from the top edge of the chassis to the hand palm is $3$~cm, we estimate that in practice the maximum uncertainty in placing the chassis within the rectangular hand palm is $\pm 2$~mm. Eight cases of 2-mm shifts toward different directions were simulated, showing an average difference up to $0.37$~dB with $0.29$~dB standard deviation. We can conclude that possible variations in dielectric properties and chassis placement on the hand palm cause up to $0.6$~dB changes in the CDF of spherical coverage compared to the original case. 

The phase center of the antenna array could not be aligned with the two rotational axes of the measurement tower, as shown in the Fig.~\ref{fig:meas_setup}(b),  due to limited adjustability on the tower. The phase center of antenna array is at $0.3$~m distance from the azimuth rotational axis and $0.36$~m from the elevation rotation axis, while the distance between the probe antenna and tower rotational axis is $6.1$~m. Due to this geometry, the rotation of the body phantom causes a maximum $\pm4^\circ$ variation in the angle in which the probe antenna sees the AUT. Due to this angular variation, the probe antenna gain towards the direction of the AUT decreases by up to $0.6$~dB, according to the specification of the probe antenna. This causes an amplitude uncertainty up to $0.6$~dB in the measurement results. Moreover, the measurement tower shadows the angles toward legs of the body phantom where $\theta>150^\circ$ as seen in Fig.~\ref{fig:realized_gain}(b).

\subsection{Results and Discussions}

\begin{figure}[t]
	\begin{center}
			\subfigure[]{\includegraphics[scale=0.5]{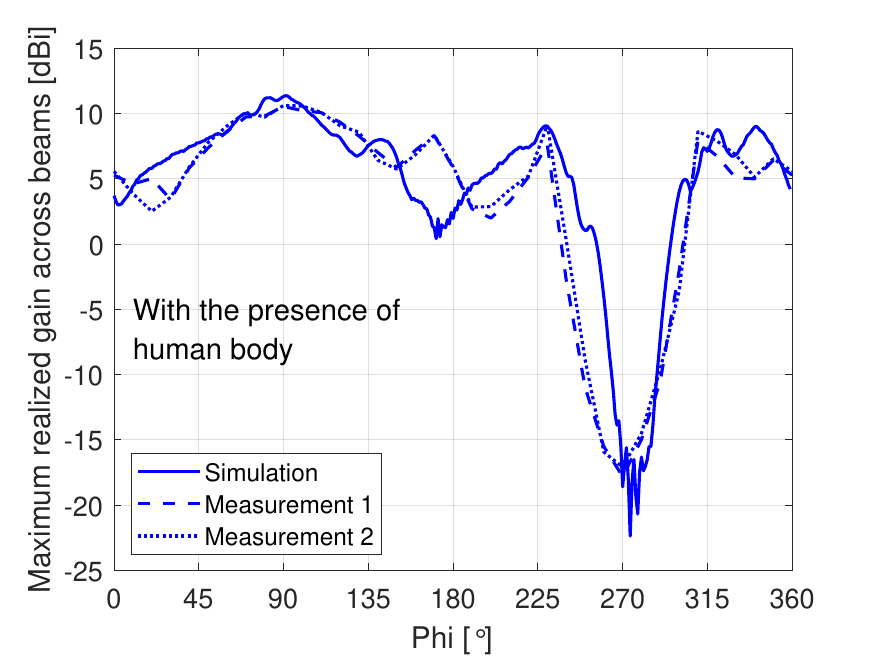}}\\
			\subfigure[]{\includegraphics[scale=0.5]{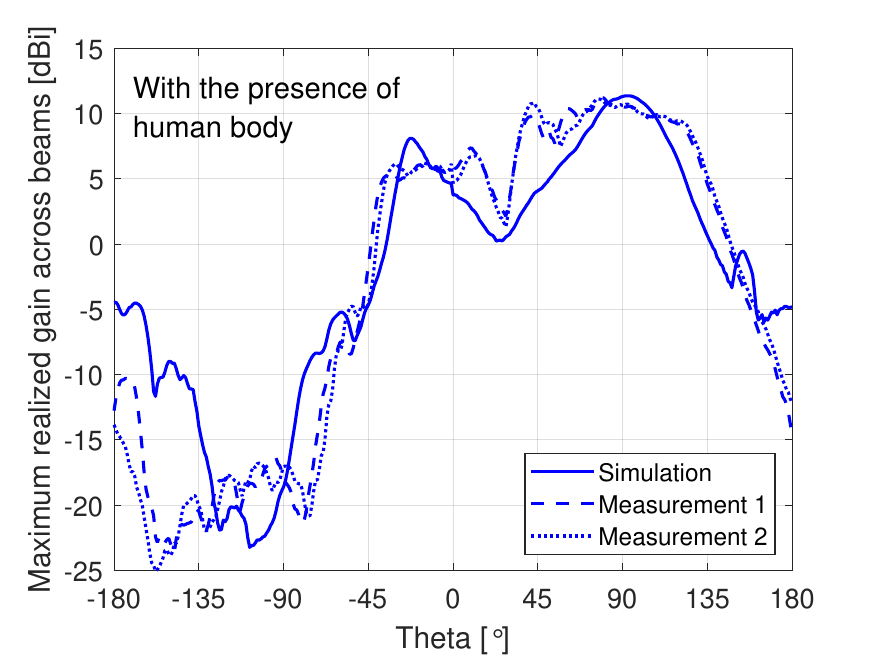}}
		\caption{Comparison of measured and simulated maximum realized gains over 2048 synthesized patterns; (a) $\theta=90^\circ$ cut, (b) $\phi=90^\circ$ cut, measurements under the presence of the hexagonal body phantom and simulations with a full numerical human model.}
		\label{fig:realized_gain}
	\end{center}
\end{figure}

The maximum realized gains of the array after pattern synthesis are illustrated in Fig.~\ref{fig:realized_gain} under the presence of the numerical human model in simulations and the hexagonal body phantom in measurements. There are a few differences between measurements and simulations. The area shadowed by the body phantom, seen in the Fig.~\ref{fig:realized_gain}(a) around  $\phi = 270^\circ$, is roughly $7^\circ$ wider in the measurements compared to the simulations. This can be caused by a difference in the width of the body phantom compared to the numerical human model. Also, the realized gain in Fig.~\ref{fig:realized_gain}(a) around $\phi=90^\circ$, and in Fig.~\ref{fig:realized_gain}(b) around $\theta=135^\circ$ is lower in the measurements than in simulations. This may be attributed to the antenna misplacement in the measurements explained in the previous paragraph. Finally, the measured realized gains near $\theta=\pm 180^\circ$ in Fig.~\ref{fig:realized_gain}(b) are lower than in the simulations due to the measurement tower shadowing those angular ranges. Although some differences can be seen from there cuts, the measured patterns match well in general with the simulated ones

\begin{figure}[t]
	\begin{center}
		\subfigure[]{\includegraphics[scale=0.5]{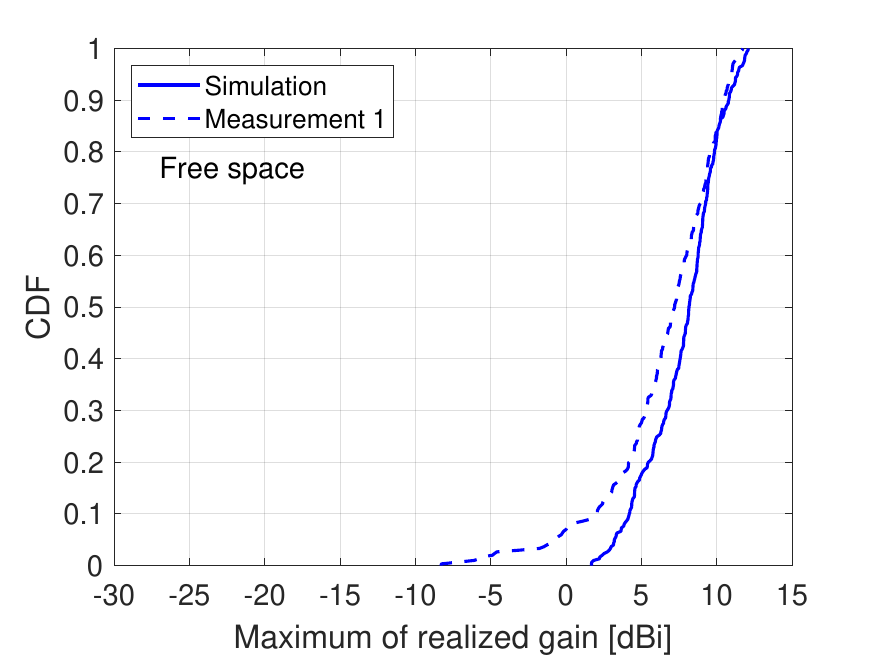}
			\label{fig:spherical_coverage_CA}}\\
		\subfigure[]{\includegraphics[scale=0.5]{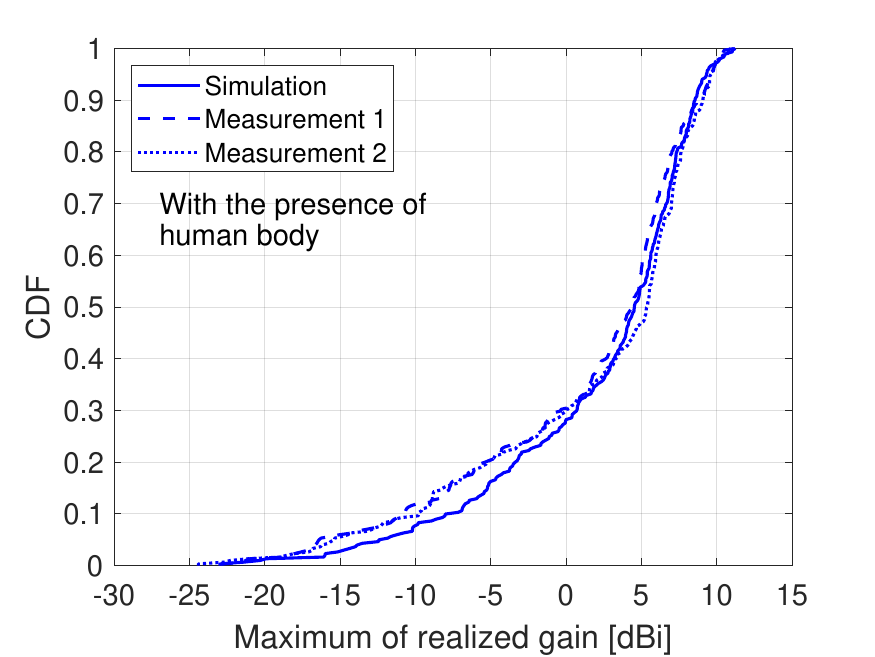}
			\label{fig:spherical_coverage_DA}}
		\caption{Statistics of spherical coverage (a) in free space and (b) with the hexagonal body phantom (measurements) and the full numerical human model (simulations).
		}
		\label{fig:spherical_coverage}
	\end{center}
\end{figure}

\subsubsection{Free space} 
The CDF of the maximum realized gains in free space across $301$ angles covering the entire sphere was calculated according to~\cite{Cziezerski19_APC} as shown in Fig.~\ref{fig:spherical_coverage}(a). It shows that the measured spherical coverage CDF of the array matches with simulated one above $0.7$ probability level. Below this level, the measured starts to get smaller gain values and at $0.1$ probability level there is $2.2$~dB difference between measured and simulated results. Differences in the outage level can be explained by the fact that absorbers were added, at the bottom part of the phone chassis, during the free space measurements to eliminate the leakage and radiation from the connectors. These absorbers shadows part of the sphere analyzed in the measurement. On the other hand, simulation do not include the absorber, leading to a possible difference in the outage level in spherical coverage CDF. 

\subsubsection{With the body phantom}
Fig.~\ref{fig:spherical_coverage}(b) shows that the presence of the body does not affect the maximum realized gain of the array compared to the free space case. Additionally, array realizes the median gain greater than $4.5$~dB. At $0.1$~outage level, the realized gains are below $-10$~dB corresponding to directions behind the body. The measured and simulated spherical coverage CDF differs by $0.3$~dB at the peak gain and $3$~dB at the $0.1$~level. Finally, the measurement with the body phantom was performed twice to verify its repeatability. There is less than $1$~dB difference between these two measurements across the CDF, which indicates good repeatability. Given the agreement between simulations using numerical human model and measurements using the hexagonal cross-sectioned body phantom, it is possible to conclude that the phantom is appropriate for evaluating antenna arrays under influence of a human body in terms of spherical coverage statistics.

\section{Conclusion}
This paper presented the design and manufacturing of a hexagonal human body phantom for repeatable tests of mobile phone antenna arrays at 28 GHz. Relevant test array was implemented on a mobile phone sized chassis and gains of it were evaluated in free space and when it was held by a hand palm of a body phantom. Measured and simulated maximum gain matches well, according to the spherical coverage CDF as an evaluation metric. We thereby demonstrated suitability of our antenna model in numerical simulations and physical body phantom design that has a simplified shape compared to actual human bodies. Proper identification of relevant body parts and shapes and the thickness of the skin material was important in order to allow straightforward manufacturing of a working body phantom. Furthermore, proper de-embedding of feed line losses was essential in comparing measured and simulated gains of fabricated antenna array. Finally, the repeatability of the antenna array measurements was demonstrated under the presence of the hexagonal human body phantom. As future work, we plan OTA measurements of various array configurations with the proposed phantom.

\bibliographystyle{IEEEtran}% bib style
\bibliography{v18}% your bib database

\begin{thebibliography}{10}
\providecommand{\url}[1]{#1}
\csname url@samestyle\endcsname
\providecommand{\newblock}{\relax}
\providecommand{\bibinfo}[2]{#2}
\providecommand{\BIBentrySTDinterwordspacing}{\spaceskip=0pt\relax}
\providecommand{\BIBentryALTinterwordstretchfactor}{4}
\providecommand{\BIBentryALTinterwordspacing}{\spaceskip=\fontdimen2\font plus
\BIBentryALTinterwordstretchfactor\fontdimen3\font minus
  \fontdimen4\font\relax}
\providecommand{\BIBforeignlanguage}[2]{{%
\expandafter\ifx\csname l@#1\endcsname\relax
\typeout{** WARNING: IEEEtran.bst: No hyphenation pattern has been}%
\typeout{** loaded for the language `#1'. Using the pattern for}%
\typeout{** the default language instead.}%
\else
\language=\csname l@#1\endcsname
\fi
#2}}
\providecommand{\BIBdecl}{\relax}
\BIBdecl

\bibitem{TR38810}
{3GPP, TR 38.810 (V16.1.0)}, ``Study on test methods (release 16),''
  http://www.3gpp.org/dynareport/38810.htm, Dec. 2018.

\bibitem{Halender16_AWPL}
J.~Helander, K.~Zhao, Z.~Ying, and D.~Sj\"{o}berg, ``Performance analysis of
  millimeter-wave phased array antennas in cellular handsets,'' \emph{IEEE
  Antennas Wireless Propag. Lett.}, vol.~15, pp. 504--507, 2016.

\bibitem{Syrytsin17_TAP}
I.~Syrytsin, S.~Zhang, G.~F. Pedersen, K.~Zhao, T.~Bolin, and Z.~Ying,
  ``Statistical investigation of the user effects on mobile terminal antennas
  for 5{G} applications,'' \emph{IEEE Trans. Ant. Prop.}, vol.~65, no.~12, pp.
  6596--6605, Dec. 2017.

\bibitem{Zhao17_AWPL}
K.~Zhao, J.~Helander, D.~Sj\"{o}berg, S.~He, T.~Bolin, and Z.~Ying, ``User body
  effect on phased array in user equipment for the 5{G mmWave} communication
  system,'' \emph{IEEE Ant. Wireless Prop. Lett.}, vol.~16, pp. 864--867, 2017.

\bibitem{Hejselbaek17_TAP}
J.~Hejselbaek, J.~{\O{}}. Nielsen, W.~Fan, and G.~Pedersen, ``Measured 21.5
  {GHz} indoor channels with user-held handset antenna array,'' \emph{IEEE
  Trans. Ant. Prop.}, vol.~65, no.~12, pp. 6574--6583, Dec. 2017.

\bibitem{Zhao17_TAP}
K.~Zhao, C.~Gustafson, Q.~Liao, S.~Zhang, T.~Bolin, Z.~Ying, and S.~He,
  ``Channel characteristics and user body effects in an outdoor urban scenario
  at 15 and 28 {GHz},'' \emph{IEEE Trans. Ant. Prop.}, vol.~65, no.~12, pp.
  6534--6548, Dec. 2017.

\bibitem{Hong17_TAP}
W.~{Hong}, K.~{Baek}, and S.~{Ko}, ``Millimeter-wave {5G} antennas for
  smartphones: overview and experimental demonstration,'' \emph{IEEE Trans.
  Ant. Prop.}, vol.~65, no.~12, pp. 6250--6261, Dec. 2017.

\bibitem{Yu18_TAP}
B.~{Yu}, K.~{Yang}, C.~{Sim}, and G.~{Yang}, ``A novel 28 {GHz} beam steering
  array for {5G} mobile device with metallic casing application,'' \emph{IEEE
  Trans. Ant. Prop.}, vol.~66, no.~1, pp. 462--466, Jan. 2018.

\bibitem{Haneda18_VTCS}
K.~Haneda, M.~Heino, and J.~J\"{a}rvel\"{a}inen, ``Total array gains of
  millimeter-wave mobile phone antennas under practical conditions,'' in
  \emph{Proc. IEEE 87th Veh. Tech. Conf. (VTC Spring 2018)}, Porto, Portugal,
  June 2018, pp. 1--6.

\bibitem{Haneda18_EuCNC}
------, ``Total array gains of polarized millimeter-wave mobile phone
  antennas,'' in \emph{Proc. European Conf. Network. Commun. (EuCNC 2018)},
  Ljubljana, Slovenia, June 2018, pp. 167--171.

\bibitem{Xu18_IEEEAccess}
B.~{Xu}, Z.~{Ying}, L.~{Scialacqua}, A.~{Scannavini}, L.~J. {Foged},
  T.~{Bolin}, K.~{Zhao}, S.~{He}, and M.~{Gustafsson}, ``Radiation performance
  analysis of 28 {GHz} antennas integrated in {5G} mobile terminal housing,''
  \emph{IEEE Access}, vol.~6, pp. 48\,088--48\,101, 2018.

\bibitem{Syrytsin18_TAP}
I.~{Syrytsin}, S.~{Zhang}, G.~F. {Pedersen}, and Z.~{Ying}, ``User effects on
  the circular polarization of 5g mobile terminal antennas,'' \emph{IEEE Trans.
  Ant. Prop.}, vol.~66, no.~9, pp. 4906--4911, Sep. 2018.

\bibitem{Raghavan19_ComMag}
V.~{Raghavan}, V.~{Podshivalov}, J.~{Hulten}, M.~A. {Tassoudji}, A.~{Sampath},
  O.~H. {Koymen}, and J.~{Li}, ``Spatio-temporal impact of hand and body
  blockage for millimeter-wave user equipment design at 28 {GHz},'' \emph{IEEE
  Commun. Mag.}, vol.~56, no.~12, pp. 46--52, Dec. 2018.

\bibitem{Raghavan19_TC}
V.~{Raghavan}, M.~{Chi}, M.~A. {Tassoudji}, O.~H. {Koymen}, and J.~{Li},
  ``Antenna placement and performance tradeoffs with hand blockage in
  millimeter wave systems,'' \emph{IEEE Trans. Commun.}, vol.~67, no.~4, pp.
  3082--3096, Apr. 2019.

\bibitem{Hazmi19_EuCAP}
A.~Hazmi, R.~Tian, S.~Rintamaki, Z.~Milosavljevic, J.~Ilvonen,
  J.~vanWonterghem, A.~Khripkov, and T.~Kamyshev, ``Spherical coverage
  characterization of millimeter wave antenna arrays in {5G} mobile
  terminals,'' in \emph{Proc. 13th European Conf. Ant. Prop. (EuCAP 2019},
  Krakow, Poland, Apr. 2019, pp. 1--5.

\bibitem{Zhao19_IEEEAccess}
K.~{Zhao}, S.~{Zhang}, Z.~{Ho}, O.~{Zander}, T.~{Bolin}, Z.~{Ying}, and G.~F.
  {Pedersen}, ``Spherical coverage characterization of {5G} millimeter wave
  user equipment with {3GPP} specifications,'' \emph{IEEE Access}, vol.~7, pp.
  4442--4452, 2019.

\bibitem{Heino19_EuCAP}
M.~Heino, C.~Icheln, and K.~Haneda, ``Self-user shadowing effects of
  millimeter-wave mobile phone antennas in a browsing mode,'' in \emph{Proc.
  13th European Conf. Ant. Prop. (EuCAP)}, Krakow, Poland, Apr. 2019, pp. 1--5.

\bibitem{Cziezerski19_APC}
C.~Cziezerski, M.~Heino, P.~Koivum\"{a}ki, K.~Haneda, C.~Icheln, A.~Hazmi, and
  R.~Tian, ``Comparing gains of 28 {GHz} module-based phased antenna arrays on
  a {5G} mobile phone,'' in \emph{Proc. 1st IET Ant. Prop. Conf. (APC 2019)},
  Birmingham, UK, Nov. 2019, pp. 1--6.

\bibitem{Syrytsin18_IEEEAccess}
I.~{Syrytsin}, S.~{Zhang}, and G.~F. {Pedersen}, ``User impact on phased and
  switch diversity arrays in 5g mobile terminals,'' \emph{IEEE Access}, vol.~6,
  pp. 1616--1623, 2018.

\bibitem{Gustafson12_RE}
C.~Gustafson and F.~Tufvesson, ``Characterization of 60 {GHz} shadowing by
  human bodies and simple phantoms,'' \emph{Radioengineering}, vol.~21, no.~4,
  pp. 979--884, Dec. 2012.

\bibitem{Aminzadeh14_EL}
R.~{Aminzadeh}, M.~{Saviz}, and A.~A. {Shishegar}, ``Theoretical and
  experimental broadband tissue-equivalent phantoms at microwave and
  millimetre-wave frequencies,'' \emph{Electronics Lett.}, vol.~50, no.~8, pp.
  618--620, Apr. 2014.

\bibitem{Lacik16_AWPL}
J.~{Lacik}, V.~{Hebelka}, J.~{Velim}, Z.~{Raida}, and J.~{Puskely}, ``Wideband
  skin-equivalent phantom for {V}- and {W}-band,'' \emph{IEEE Ant. Wireless
  Prop. Lett.}, vol.~15, pp. 211--213, 2016.

\bibitem{MakeHuman}
{Make Human}, http://www.makehumancommunity.org.

\bibitem{Gabriel97_PMB}
S.~Gabriel, R.~Lau, and C.~Gabriel, ``The dielectric properties of biological
  tissues: {III}. parametric models for the dielectric spectrum of tissues,''
  \emph{Physics in medicine and biology}, vol.~41, pp. 2271--93, Dec. 1996.

\bibitem{Wu15_MM}
T.~{Wu}, T.~S. {Rappaport}, and C.~M. {Collins}, ``Safe for generations to
  come: considerations of safety for millimeter waves in wireless
  communications,'' \emph{IEEE Microwave Magazine}, vol.~16, no.~2, pp. 65--84,
  Mar. 2015.

\bibitem{Sasaki14}
K.~Sasaki, K.~Wake, and S.~Watanabe, ``Measurement of the dielectric properties
  of the epidermis and dermis at frequencies from 0.5 ghz to 110 ghz,''
  \emph{Phys Med Biol.}, pp. 4739--47, 8 2014.

\bibitem{Chahat12_EL}
N.~{Chahat}, M.~{Zhadobov}, S.~{Alekseev}, and R.~{Sauleau}, ``Human
  skin-equivalent phantom for on-body antenna measurements in 60 {GHz} band,''
  \emph{Electronics Letters}, vol.~48, no.~2, pp. 67--68, Jan. 2012.

\bibitem{Dancila14_SRT}
D.~Dancila, R.~Augustine, F.~Töpfer, S.~Dudorov, X.~Hu, L.~Emtestam,
  L.~Tenerz, J.~Oberhammer, and A.~Rydberg, ``Millimeter wave silicon
  micromachined waveguide probe as an aid for skin diagnosis – results of
  measurements on phantom material with varied water content,'' \emph{Skin
  Research and Technology}, vol.~20, no.~1, pp. 116--123, 2014.

\bibitem{Olkkonen13_EuCAP}
M.~{Olkkonen}, V.~{Mikhnev}, and E.~{Huuskonen-Snicker}, ``Complex permittivity
  of concrete in the frequency range 0.8 to 12 ghz,'' in \emph{2013 7th
  European Conf. on Ant. and Prop. (EuCAP)}, 2013, pp. 3319--3321.

\bibitem{Krupka16_MWCL}
J.~{Krupka}, ``Measurements of the complex permittivity of low loss polymers at
  frequency range from 5 {GHz} to 50 {GHz},'' \emph{IEEE Microwave and Wireless
  Components Letters}, vol.~26, no.~6, pp. 464--466, Jun. 2016.

\bibitem{ITU-R_P2040}
{Recommendation ITU-R P.2040}, ``Effects of building materials and structures
  on radiowave propagation above about 100 {MHz},''
  https://www.itu.int/dms\_pubrec/itu-r/rec/p/R-REC-P.2040-1-201507-I\!\!PDF-E.pdf,
  July 2015.

\bibitem{mikko_heino_2019_3249975}
M.~Heino, C.~Icheln, and K.~Haneda, ``{Simulated self-user shadowing for mobile
  phone antennas at 28 {GHz} and at 60 {GHz}},'' Jun. 2019, available at
  \url{https://doi.org/10.5281/zenodo.3249975}.

\bibitem{vaha_savo_lauri_2021_4558779}
\BIBentryALTinterwordspacing
L.~Vähä-Savo, M.~Heino, K.~Haneda, C.~Icheln, A.~Hazmi, and R.~Tian,
  ``{Numerical and experimental body phantoms for simulating and measuring
  radiation patterns of a mobile phone at 28 GHz},'' Feb. 2021. [Online].
  Available: \url{https://doi.org/10.5281/zenodo.4558779}
\BIBentrySTDinterwordspacing

\bibitem{Molisch10_book}
A.~Molisch, \emph{Wireless communications, 2nd Edition}.\hskip 1em plus 0.5em
  minus 0.4em\relax Wiley, 2010.

\bibitem{R4-1700095}
3rd Generation Partnership Project~({3GPP}), ``Discussion of mmwave {UE} {EIRP}
  and {EIS} test,'' 3GPP TSG-RAN WG4 NR AH Meeting R4-1700095, Jan. 2017,
  available at
  \url{https://www.3gpp.org/ftp/tsg_ran/WG4_Radio/TSGR4_AHs/TSGR4_NR_Jan2017/Docs/R4-1700095.zip}.

\end{thebibliography}

\end{document}